# Excited state spectroscopy and spin splitting in atomically thin quantum dots

P. Kumar[1,2], H. Kim[4], S. Tripathy[4], K. Watanabe[5], T. Taniguchi[5], K. S. Novoselov[1,2,3*], D. Kotekar-Patil[4*†]

1. Institute for Functional Intelligent Materials, National University of Singapore, Singapore, 117544 Singapore
2. Integrative Sciences and Engineering Programme, National University of Singapore, 119077 Singapore
3. Department of Materials Science and Engineering, National University of Singapore, Singapore, 117575 Singapore
4. Institute of Materials Research and Engineering, A∗STAR (Agency for Science, Technology and Research), Innovis, 2 Fusionopolis way, Singapore, Singapore 117602, Singapore
5. Research Center for Functional Materials, National Institute for Materials Science, Tsukuba, 305-0044 Japan

*__Email:__ dharmraj_kotekar_patil@imre.a-star.edu.sg, kostya@nus.edu.sg

**Semiconducting transition metal dichalcogenides (TMDCs) are very promising materials for quantum dots and spin-qubit implementation. Reliable operation of spin qubits requires the knowledge of *Landé g*-factor, which can be measured by exploiting the discrete energy spectrum on a quantum dot. However, the quantum dots realized in TMDCs has yet to reach the required quality for reliable measurement of g-factor. Quantum dot sizes reported in TMDCs so far are not small enough to observe discrete energy levels on them. Here, we report on electron transport through discrete energy levels of quantum dot in a single layer $MoS_2$. The quantum dot energy levels are separated by few (5-6) meV such that the ground state and the excited state transitions are clearly visible. This well resolved energy separation allows us to accurately measure the ground state *g*-factor of ~5 in $MoS_2$ quantum dots. We observe a spin filling sequence in our quantum dot under perpendicular magnetic field. Such a system offers an excellent testbed to measure the key parameters for evaluation and implementation of spin-valley qubits in TMDCs, thus accelerating the development of quantum systems in two dimensional semiconducting TMDCs.**

**Introduction:**

Two-dimensional semiconducting transition metal dichalcogenides (TMDCs) offer versatile properties with applications in a wide range of areas including microelectronic, optoelectronics and quantum devices. The crystal structure with two different atoms in its hexagonal lattice - gives rise to two inequivalent valleys (namely *K* and *K'*) in the Brillouin zone. Moreover, TMDCs usually exhibit strong spin orbit coupling (SOC) which originates from the heavy metal atom and, as a result, provides coupling between the spin and the inequivalent valleys. This coupling between the spin and valley is predicted to enhance the coherence time significantly.[1] SOC has been shown to allow fast electric-dipole spin manipulation,[2,3] making them an attractive material platform for implementation of spin-valley quantum bits (qubits).

Spin manipulation in spin (-valley) qubits relies on exciting the electron by microwave frequency and matching it to the Zeeman splitting. This requires precise information on the *Landé g*-factor. *Landé g*-factor can be precisely measured from the Zeeman splitting of the quantum dots (QD) energy levels. For such a measurement, a QD with its energy level separation ($\Delta E$) well resolved at a given thermal energy ($k_B T$, known as the quantum regime) is required.[4] To fulfil this condition, often smaller QDs are required with relatively few charge carriers on them. Tuning the device in quantum regime has been relatively easy in advanced material systems like GaAs,[5] silicon[6,7] and germanium[8] based heterostructure which exhibit relatively low effective mass and the technology to achieve low contact resistance in these materials is well established. On the other hand, TMDCs have relatively large effective mass [9,10] and achieving transparent ohmic contacts has been challenging,[11] particularly at cryogenic temperatures. Several demonstrations of QDs in TMDCs have been reported,[12–17] however the quality of QDs achieved so far are not sufficient to precisely measure the *g*-factors. Most of the QDs reported are relatively large in size such that the individual energy level spacings becomes indistinguishable relative to the thermal energy ($\Delta E < k_B T$). Observing excited state spectroscopy in QD with similar clarity as in GaAs or silicon is still missing in TMDC, resulting in the lack of basic understanding of material properties in the quantum regime where spin-valley qubits are operated. There are several factors that limit such a detailed characterization of QD. First, transparent ohmic contact to a QD is necessary so that the QD spectroscopy can be performed without being impeded by noise due to low current flowing through the device or requirement of large gate voltages which pushes the QD in the metallic regime. Second, the QD should be small enough with relatively few charge carriers on it so that their energy level spacings are not blurred by the thermal energy. Achieving both these conditions could facilitate detailed characterization of QDs in TMDCs.

Here, we report on single electron transport through a single layer MoS$_2$ QD with low contact resistance and high mobility which enables us to probe the excited states on the QDs. The device operates at relatively low gate voltages, restricting it to a few charge carriers on the QDs. Using a dual gated geometry, we leverage QDs originating intrinsically in the sample, e.g. vacancies or sample inhomogeneity (e.g. nanobubbles) which are typically smaller in size (<100 nm). [18–24] In such a device, we study resonant tunneling of electrons through QDs in a single layer MoS$_2$. In our dual gated device, the back gate accumulates electron gas in the MoS$_2$ channel whereas the top gate screening couples the QDs underneath it to the electron gas through a tunnel barrier, thereby allowing to perform tunnel spectroscopy. We perform excited state spectroscopy and study evolution of these QD energy levels under perpendicular magnetic field. We show that the QDs are strongly coupled to the top gate and therefore can be tuned by the top gate, which opens the possibility of electrically controlling the individual spins.[25,26]

To perform tunnelling spectroscopy through the QDs in TMDC, we prepared a vdW heterostructure where a single layer $MoS_2$ was contacted with a few-layer graphite (FLG) and encapsulated between two hexagonal boron nitride (hBN) flakes as shown schematically in Figure 1A. The stack was assembled inside $Ar/N_2$ filled glovebox with oxygen concentration < 1ppm. Subsequently the stack was annealed at 300°C for 1 hr in forming gas. Cr/Au electrodes were used as one-dimensional edge contacts to FLG and a narrow top gate was placed on $MoS_2$ channel, schematically shown in Figure 1B.[27] Optical image of the fabricated device is shown in Figure 1C. We validated the single layer nature of $MoS_2$ by a photoluminescence (PL) measurement exhibiting a strong PL peak centered at 1.86eV at room temperature using 514nm excitation (Figure 1D). Our sample employs a dual gated geometry where a global back gate controlling the carrier density in the entire $MoS_2$ channel and a narrow top-gate (gate length = 100 nm), locally controlling the carrier density in the tunnel barriers. We target to use a relatively thin hBN (~ 10 nm) for top gate dielectric which was initially chosen by optical contrast and later confirmed by TEM cross-section (Figure 1E). TEM cross-section image shown in Figure 1E is taken on another sample, prepared by following the same process as the device measured in this work. This thin hBN brings the top gate closer to the $MoS_2$ channel. Consequently, the top gate partially screens the applied back gate voltage. When positive back gate voltage is applied, it accumulates electrons non-uniformly along the channel, with higher carrier concentration outside the top gated region and lower carrier concentration underneath the top gate due to screening. This creates a barrier shaped electrochemical potential profile along the length of the channel as shown in Figure 1F (red profile). Top gate screening allows us to separate the QD from the electron reservoir by a tunnel barrier (Figure 1F). Only the QDs underneath the top gate participate in the resonant tunneling whereas the other defects outside the top gated region are completely occupied by electrons, due to doping by the global gate.

We characterized our device by applying a source-drain voltage ($V$) and measure the drain current ($I_d$) through the device. First, we selected the range of the back gate voltages ($V_{bg}$) to ensure the low contact resistance for our QDs. To this end we measure two-point resistance of the FLG on the same side of the device (Figure 1G) as well as across the device at two different top gate voltages ($V_{tg}$) (Figure 1H and 1I). Two-point resistance across FLG at $V_{bg} = 30V$ gives us $2R_{C,FLG} \leq 260\Omega$ (Figure 1G) which is comparable to the expected value of one dimensional edge contact.[27] Furthermore, we measure the two-point resistance across the device by applying $V_{tg} = 0V$ and $V_{tg} = 4V$. When $V_{tg} = 0V$, with increasing $V_{bg}$, $I_d$ exhibits oscillatory behavior and devices remain in highly resistive state (200 K$\Omega$) even at $V_{bg} = 70V$ (Figure 1H). When $V_{tg} = 4V$, device current starts to increase rapidly above $V_{bg} \geq 5V$ switching from an insulating state at $V_{bg} \leq 5V$ to a metallic state ($\leq 2k\Omega$) at $V_{bg} \geq 7V$ (Figure 1I). Low pinch-off voltage (~ 5V) with rapid transition from OFF-state to ON-state despite back gate placed ~ 300nm away from the channel, indicates a negligible Schottky barrier at the FLG and single layer $MoS_2$ interface. Moreover, $I_d$-$V_{bg}$ measurements at two different $V_{tg}$ indicates that the device resistance in Figure 1H and 1I is mainly dominated by the $MoS_2$ segment underneath the top gate while the $MoS_2$ segment outside the top gate enters a metallic state above $V_{bg} = 7V$. We get a total resistance across the device ($R_{tot}$) of ~ 1.4k$\Omega$ at $V_{tg} = 4V$ and $V_{bg} = 20V$. This sets an upper limit on the contact resistance of $2R_c \leq 1.4k\Omega$. Thus, we have achieved low contact resistance with high electron mobility (Figure S0, supplementary section) and ohmic contact is not the bottleneck in our device performance.

## Figure 1: Device fabrication and characterization

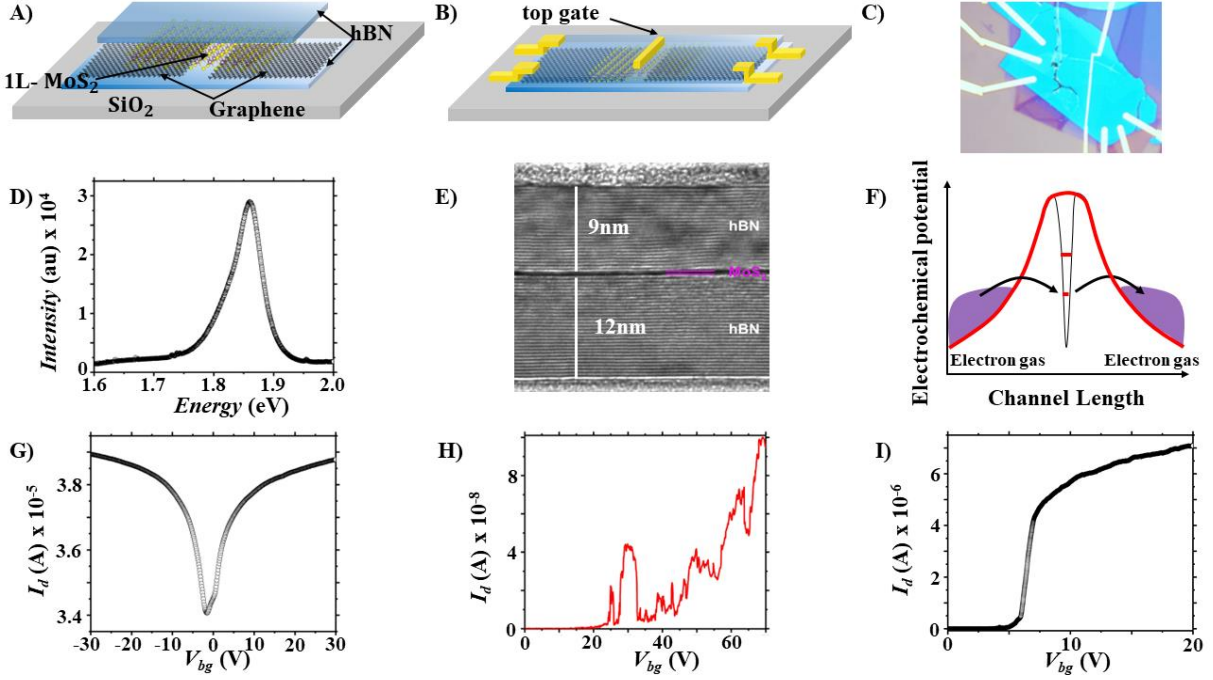

A) Schematic of the MoS$_2$-based van der Waals heterostructure stack with single layer MoS$_2$ using few layers graphene contacts encapsulated between two hBN flakes. B) Schematic of the van der Waals heterostructure device. The device uses few layers graphene to contact single layer MoS$_2$. Cr/Au edge contacts to few layer graphene are defined using e-beam lithography, reactive ion etching and thermal evaporator. A top gate (Ti/Au) of width 100nm is patterned to control local electron density in the MoS$_2$ channel. C) Optical microscope image of the device measured in this work. D) Room temperature photoluminescence spectrum for MoS$_2$ heterostructure. The PL peak is centred at ~1.86 eV consistent with single layer MoS$_2$ bandgap. We use a 514nm excitation laser. E) Cross section TEM image of the hBN/MoS$_2$/hBN vdW heterostructure showing hBN layers close to 10nm thickness and 0.7 nm single layer MoS$_2$. F) Electrochemical potential profile exhibiting a tunnel barrier profile along the MoS$_2$ channel due to screening from the top gate (red profile). QD potential (black profile) underneath the top gate adds localized states allowing resonant tunnelling studies through the device. G) Two-point resistance between few layer graphene measured on the same side of the device measured by applying $V=10mV$. H) and I) $I_d$-$V_{bg}$ measured at $V=10mV$, $V_{tg}=0V$ and $V_{tg}=4V$ respectively.

## Results:

We first look at the transport regime where $V_{tg} = 0V$. The MoS$_2$ channel segment underneath the top gate dominates the device resistance. Figure 2A shows line traces of $G$ ($I_d$/$V$) vs $V_{bg}$ at $V_{tg} = 0V$ measured at $T = 1K$. We will focus on the range of back gate voltages above 7V, where our two probe measurements demonstrate low contact resistance. In this regime the contact resistance to the quantum dot varies negligibly (Figure 1I), and the back gate mainly affects the states inside the quantum dots. In this regime we observe resonant tunneling quasi-periodic G oscillations reminiscent to Coulomb blockade (CB) oscillations due to single electron tunneling. To further confirm, we measure a differential conductance ($dI_d$/$dV$) color map as a function of $V$ and $V_{bg}$ as shown in Figure 2B. The $G$ oscillations in Figure 2A evolve into diamond shaped domains in the $V$-$V_{bg}$ space where $dI_d$/$dV$ is strongly suppressed inside the diamond and increases in steps outside the diamond. This characteristic is a hallmark of CB diamonds due to a QD coupled to its environment through a tunnel barrier resulting in resonant tunneling of individual electrons. Multiple CB diamonds are observed. We note that the CB diamonds measured in our device are free of charge noise, thanks to highly transparent ohmic contacts. Here, we study three of the CB diamonds in detail (marked with A, B and C). Over the whole color map in Figure 2B, we note an asymmetry in $dI_d/dV$ amplitude where the amplitude is larger for the positive bias compared to the negative bias. This asymmetry is a consequence of asymmetric tunnel coupling to the source and drain regions.[24] In the $V_{bg}$ range 30V - 40V, no features are observed inside the CB diamond for positive bias voltage due to saturation of current amplifier.

**Figure 2: Quantum dot spectroscopy ($T$ = 1K)**

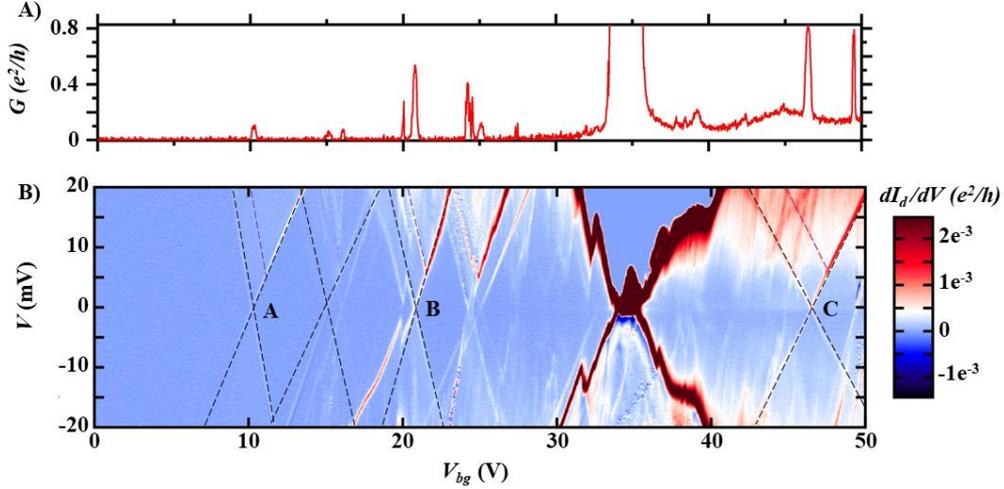

*A) Conductance trace measured at T = 1K with a source-drain bias V = 1mV applied. Sharp conductance resonances observed are associated with single electron tunnelling events through the device. B) Differential conductance bias spectroscopy in the V-$V_{bg}$ space exhibiting multiple Coulomb diamonds marked with black dashed lines. Differential conductance resonances running parallel to the ground state of Coulomb diamond A, B and C are attributed to the excited states of the QDs and are marked with purple dashed lines.*

Figure 3A shows the zoomed in region of diamond A (marked in Figure 2B). The charging energy ($E_C$) of the QD can be determined by the height of the CB diamond along the V-axis which is ~ 19meV (Figure 3A). QDs capacitive coupling to its environment is reflected through the CB diamonds slopes which give a total capacitance ($C_{tot} = C_s + C_d + C_g$) of 8aF, where $C_s$ (5.47aF), $C_d$ (2.54aF) and $C_g$ (0.03aF) is the source, drain and gate capacitance respectively). Charging energy ($e^2/C_{tot}$ = 19.8meV) estimated using the extracted capacitances from the CB diamonds agrees well with the measured charging energy (19meV, Figure 3A). We estimate the QD size by assuming it to be a disk shaped, $E_c = e^2/8\varepsilon_0\varepsilon_r r$, which gives $r$ ~ 29 nm. The QD size of a few tens of nanometers suggests that the origin of quantum dot is unlikely to be point defects in $MoS_2$. We suspect that the origin of QD could be the sample inhomogeneity.

For a small QD with relatively few charge carriers (electrons) in them, the energy level spacing on the QD is large enough to overcome thermal broadening and can be resolved in transport measurements. This visibility of excited state indicates that the QD is in the quantum limit, one of the key requirements for the demonstration of qubits. In Figure 3a, in addition to the ground state transition (CB diamond crossing, black dashed lines), we observe additional resonance running parallel to the ground state. The resonance intersects the ground state at a finite bias. This resonance is attributed to the excited state on the QD which lies 5meV above the ground state.[25,28] Similar resonances are also observed for CB diamond B and C which lies 5meV and 6meV above the ground states respectively, indicating that our QDs are in the quantum regime.

In the quantum regime, electron transport takes place through a single energy level of the QD as the energy levels are well separated with respect to thermal broadening under small bias voltage applied. A standard method to distinguish between single and multi-level transport in a QD is to measure the temperature dependence of the CB peak amplitude. The peak resistance scales linear in temperature in the single-level case whereas in the multi-level case the peak resistance is temperature independent.[6,29,30] In Figure 3B, we show the CB peak evolution for temperatures ranging from 1 to 3.4K. The CB peak resistance is plotted in Figure 3C as a function of temperature. The linear dependence of CB peak resistance with temperature suggests single level transport. This further validates our claim that our device is operating in quantum limit i.e. $\Delta E > k_B T$. This is consistent with the visibility of QD excited state in our device.

**Figure 3: Excited state spectroscopy and single level transport**

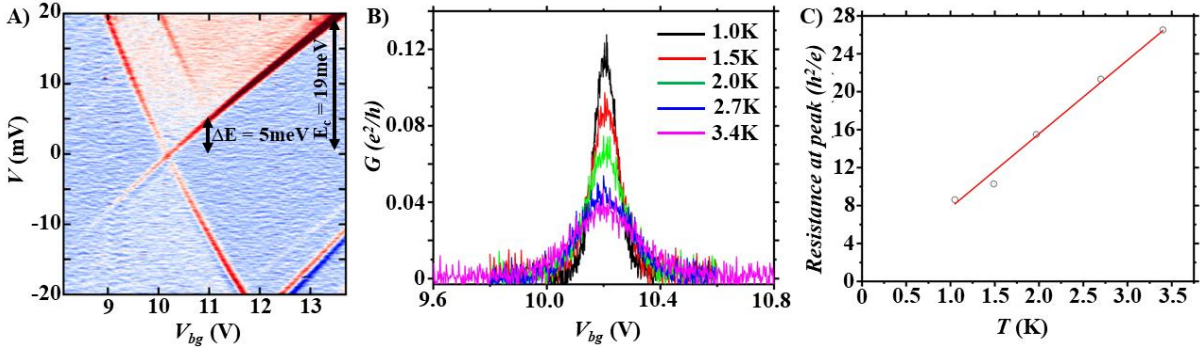

*A) Zoomed in region of CB diamond A shown in figure 2B. Excited state on the QD is seen as a differential conductance resonance running parallel to the ground state, which lies 5meV above the ground state. From the size of the Coulomb diamond a charging energy of 19meV is measured. B) Temperature dependence of the CB peak is studied at various temperatures. With increasing temperature, a drop in conductance in recorded. C) Peak resistance vs temperate exhibits a linear dependence consistent with single level transport regime.*

We next look at the evolution of these individual QD states under perpendicular magnetic field. External magnetic field lifts the spin degeneracy, e.g. spin-up forming the ground state and spin-down forming the excited state where the separation between them is given by Zeeman energy ($E_z = g\mu_B B$) as schematically shown in Figure 4A (where $g$ is the *Landé g*-factor, $\mu_B$ is the Bohr's magneton, $B$ is the external magnetic field). Energetically accessible electron transitions are only visible in the V-shaped region of the CB diamond as shown in Figure 4B and 4C (orange region) and transitions are blocked inside the CB diamonds (yellow region). In the presence of Zeeman splitting, differential conductance lines inside the V-shaped region indicate the transitions involving the ground and excited spin states. If the electron entering the QD occupies the spin-up ground state, the spin-down excited state appears as a differential conductance line terminating at the edge of *N+1* electron CB region (Figure 4B). However, if the electron entering the QD occupies the spin-down excited state, the spin-up ground state appears as a differential conductance line terminating at the edge of *N* electron CB region (Figure 4C).

Spin filling information can be inferred from the splitting observed in the CB diamonds under external magnetic field. Additionally, it can also provide a direct measure of the *g*-factor in the QD. For this, we compare the same CB diamond under different magnetic field to observe the evolution of QD states. Figure 5A-D shows CB diamond B measured for $B_z = 0$T, 2.5T, 5T and 7.5T. We observe splitting of the ground state marked by the black arrows in Figure 5A-D.[31] In Figure 5E we plot the extracted Zeeman energy as a function of the $B_z$. The Zeeman energy follows a linear dependence. The slope of this line can be fitted using $E_z/\mu_B B = g$. A linear fit intersecting at $B_z = 0$T gives us a g-factor of $5.20 \pm 0.17$ in our MoS$_2$ QD. Similar measurements have been performed on CB diamond A and C. While g-factor extracted on Diamond C (Figure S2, $g = 5.15 \pm 0.23$) matches well with that for diamond B, diamond A shows lower g-factor of 3.5 (Figure S1). Comparing the measured CB diamonds under finite magnetic field with the schematic shown in Figure 4B and 4C, we note that the electrons prefer to enter the ground state first before occupying the excited state. Similar trend is observed for all the three CB diamonds A, B and C. This picture is consistent when an electron is occupying an empty orbital in the QD suggesting that our QD is either completely empty to begin with or occupying even number of electrons.[5] We note that the g-factor values measured in our QD device is larger than the previously reported g-factor in single-layer MoS$_2$ device.[32] However, due to the large effective electron mass in MoS$_2$ (and TMDC in general), the orbital effects of magnetic field are ignored when extracting g-factors, even though they may contribute to the Zeeman shifts in magnetic field.[33–35]

**Figure 4: Zeeman splitting**

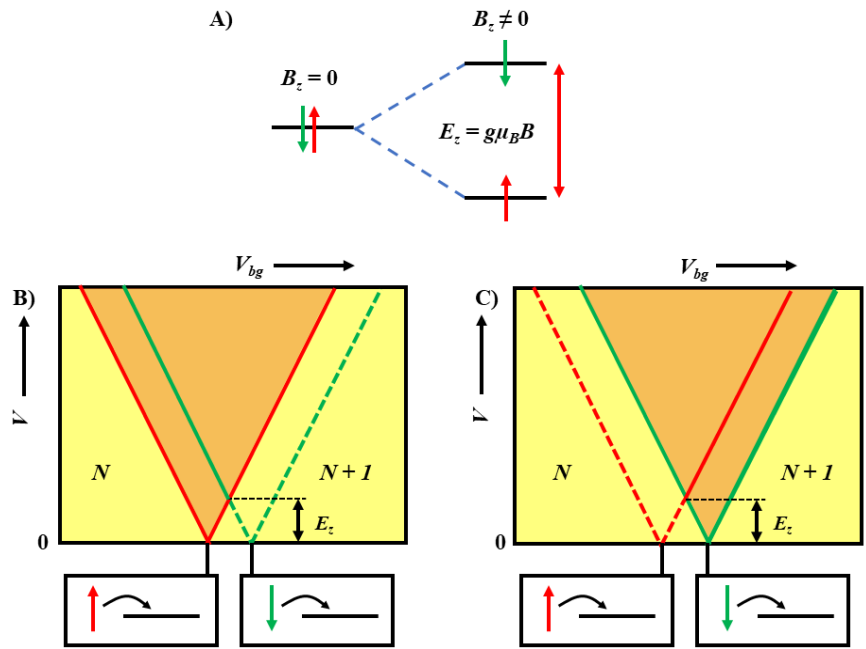

A) Schematic of the Zeeman splitting in a quantum dot. In the absence of external magnetic field, the spin states on the QD are degenerate. When external magnetic field is applied, the spin degeneracy is lifted causing splitting of the spin states. Zeeman splitting is linearly proportional to the strength of the external magnetic field and is given by $E_z = g\mu_B B$. B) Schematic of the CB diamond if electron first occupies spin-up ground state then the spin-down appears as a differential conductance line terminating at edge of N+1 electron CB region. The green dashed lines shown are in the CB region and hence are not visible in the transport measurements. C) Schematic of the CB diamond if electron first occupies a spin-down state and the spin-up appears as a differential conductance line terminating at edge of N+1 electron CB region. The red dashed lines shown are in the CB region and hence are not visible in the transport measurements.

**Figure 5: Spin filling and *g*-factor**

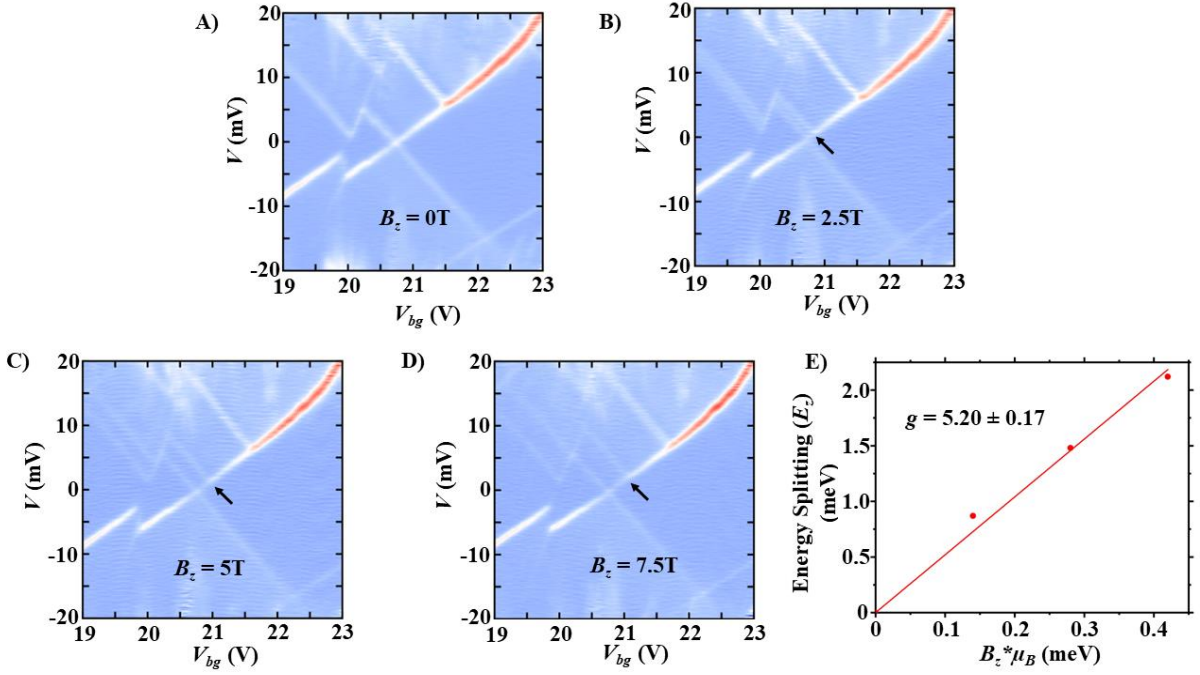

*Panel A- D shows CB diamond measured under varying magnetic field from $B_z = 0 - 7.5T$ in steps of 2.5T respectively. Splitting of the ground state is observed marked with a black arrow. E) Zeeman splitting measured from data in panel A-D is plotted as a function of magnetic energy scale. A linear fit gives the g-factor of g = 5.20±0.17.*

Lastly, we look at the coupling of the QDs to the top gate and its position relative to the top gate. To understand the top gate dependence, we perform three sets of measurements with varying $V_{tg}$. We first measure the CB diamond map at $V_{tg} = 0V$ as shown in Figure 6A. These measurements have been performed in a separate cool-down, so the CB diamonds appear to be slightly shifted at higher $V_{bg}$ (~ 59V) compared to the first cool down (Figure 2B). When we increase $V_{tg} = 4V$, CB diamonds are no longer visible suggesting confinements are completely removed (Figure 6B). A strong dependence of CB on $V_{tg}$ suggests that the CB originating from the QD is strongly coupled to the $V_{tg}$. This is in good agreement with our hypothesis that only the QDs underneath the top gate allow for resonant tunneling through its individual states. This is further validated by measuring the CB diamond map by sweeping the $V_{tg}$ at a fixed $V_{bg}$ = 59V (first Coulomb diamond position) where we observe a strong dependence of QD energy levels on $V_{tg}$ (Figure 6C). Above $V_{bg} > 7V$, MoS$_2$ channel enters metallic state (Figure 1I and 6B) and a presence of QD strongly dependent on $V_{tg}$ verifies that the QD are placed underneath the top gate supporting our claim, schematically shown in Figure 6D. In this operating regime ($V_{tg} = 0V$), the device resistance is dominated by the MoS$_2$ channel underneath the top gate. When we increase $V_{tg}$, it accumulates electrons below the top gate forming an electron gas removing all the confinement (Figure 6E). We can estimate the top gate coupling strength relative to back gate by measuring the QD charge transition in the $V_{tg}$ - $V_{bg}$ space (Figure S4). From the slope of the charge transition in Figure S4, we estimate that the top gate is ~ 69 times strongly coupled to QDs (~ 0.2aF) compared to back gate (Figure S4). The strong coupling of the top gate is a consequence of relatively thin hBN used in our device as a top gate dielectric (~ 10nm).

**Figure 6: Quantum dot position and coupling to top gate ($T$ = 160mK)**

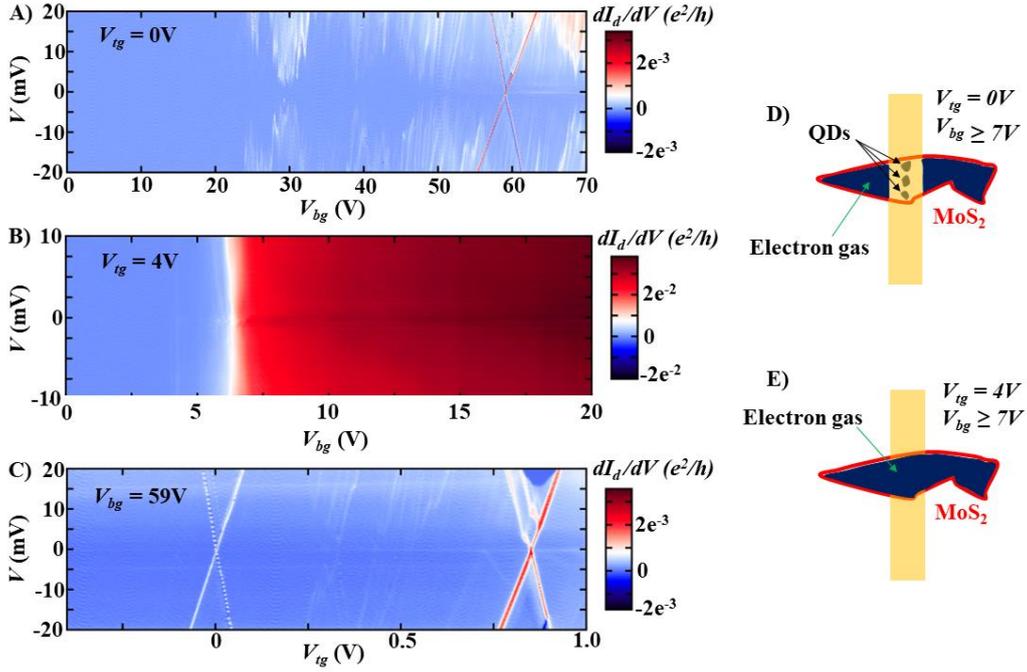

*A) Differential conductance bias spectroscopy in the V-$V_{bg}$ space while holding $V_{tg}$ = 0V. The measurement is performed on the same device shown in figure 2B in a second cool down at T = 160mK. The device switched to another charge configuration compared to the first cooldown shown in figure 2B resulting is shift of the Coulomb diamond to a higher $V_{bg}$ = 59V. B) Differential conductance bias spectroscopy performed at $V_{tg}$ = 4V. Coulomb diamonds seen in panel A are completely washed out with a sharp transition from insulating state for $V_{bg}$ <6V to metallic state for $V_{bg}$ >7V. Disappearing of Coulomb diamonds with increasing $V_{tg}$ suggest that the QDs are well coupled to the top gate and can be tuned with the top gate. C) Differential conductance bias spectroscopy in V-$V_{tg}$ space with $V_{bg}$ = 59V exhibiting CB diamonds further verifying the presence of quantum dot underneath the top gate. D and E schematically displays the channel configuration as a function of top gate voltages.*

**Discussion:**

In summary, we present resonant tunneling through QDs in single layer $MoS_2$ and study its magnetic field evolution on single particle level. We observe that our device exhibits relatively low charge noise as seen from the CB diamonds. This is a consequence of low contact resistance and high electron mobility demonstrating high van der Waals stack quality. By applying a positive back gate voltage, we accumulate electron gas in the $MoS_2$ channel while the top gate isolates the QDs underneath it, which allows us to study resonant tunneling through the QDs. We observe CB diamonds with clear excited states separated from their ground state by few meV. Performing bias spectroscopy under different magnetic fields allowed us to extract Zeeman splitting and *g*-factor values in the $MoS_2$ conduction band. CB diamond features under finite magnetic field gives insight into specific spin filling order which suggests that our QDs are completely empty to begin with or occupied with even number of electrons. Transparent ohmic contact along with clear QD spectroscopy in the quantum regime marks an important milestone for TMDC quantum devices towards implementation of spin-valley qubits. Such QD system not only allows to study intrinsic properties in $MoS_2$ but can also serve as a test bed for measuring electron coherence times in a spin-valley qubits.[36]

## Methods:
### vdW heterostructure preparation:

Prior to assembly of the stack, we mechanically exfoliated hBN, Graphene (NGS Graphite) and $MoS_2$ (HQ Graphene) flakes on highly doped Si wafers with 285 nm thick $SiO_2$ (NOVA Electronic Materials). We used polycarbonate (PC) coated PDMS stamp to assemble the stack. First, we pick-up a thin hBN flake (~10 nm) using the PC/PDMS stamp, subsequently 1L $MoS_2$ (1L-thick flake is identified by optical contrast) is picked-up by the hBN flake on the PC/PDMS stamp. Next, we pick up two pieces of few-layer graphene to make electrical contact around the perimeter of the $MoS_2$ channel. Finally, we transfer the entire stack on a bottom hBN flake (30 nm). The exfoliation and transfer of the flakes was performed inside $Ar/N_2$ filled glovebox with oxygen concentration < 1 ppm. This is a crucial step to make high-quality transport devices with low contact resistance. After fully encapsulating the stack with hBN, the sample was removed from the glovebox. Next, the sample was annealed in forming gas ($Ar/H_2$) environment at 300 °C for 1 hour to remove organic residues on the sample surface.

### Device fabrication of hBN-encapsulated dual gated $MoS_2$ device:

1D-edge contact to graphene was performed by dry etching (Oxford Plasmalab System 100) and thermal evaporation (Kurt J. Lesker Nano 36) of Cr/Au (7/60 nm). In another lithography step, top gate structure (100nm wide) was defined using electron beam lithography (EBL) followed by Cr/Au (5/25 nm) metal deposition using thermal evaporator.

### Electrical measurements:

Electrical characterization is carried out using a Nanonis Tramea DACs and a Femto DLPCA-200 current amplifier. Cryogenic measurements were performed in an Bluefors dilution refrigerator system with a base temperature of 100 mK equipped with a 9T solenoid magnet.

### Transmission electron microscopy:

Dual beam focused ion beam (FIB) technique (FEI, DB Helios 450S Nanolab) was applied to prepare an ultra-thin sample (70 nm) for transmission electron microscopy (TEM) measurements. TEM and elemental analysis were analyzed by a TEM system (FEI, Tecnai G2 TF20) with EDX analyzer.

### Optical characterization:

Room temperature PL and Raman spectra were obtained with a 532 nm laser with objective x100 (NA = 0.85) and 2400 l/mm grating.

**Acknowledgement:**

This research is supported by the Ministry of Education, Singapore, under its Research Centre of Excellence award to the Institute for Functional Intelligent Materials (I-FIM, project No. EDUNC-33-18-279-V12). KSN is grateful to the Royal Society (UK, grant number RSRP\R\190000) for support. DKP is grateful to the Agency for Science, Technology, and Research (A*STAR) for support under it's A*STAR Career Development Fund (Grant number. C210112039). K.W. and T.T. acknowledge support from JSPS KAKENHI (Grant numbers 19H05790, 20H00354 and 21H05233).

**Author contributions**

P.K prepared the vdW heterostructure and assisted D.K.P in device fabrication. H.K performed the transmission electron microscope imaging. S.T performed the optical characterization of the sample. K.W and T.T provided the hBN crystals. K.S.N and D.K.P conceived and directed the research. D.K.P performed the low temperature characterization and data analysis in discussion with K.N.S. D.K.P and K.N.S wrote the manuscript with input from all the authors.

Correspondence and requests for materials should be addressed to K. S. Novoselov (kostya@nus.edu.sg) or D. Kotekar-Patil (dharmraj_kotekar_patil@imre.a-star.edu.sg).

†**Current address: University of Arkansas, 731 West Dickson Street, Fayetteville, AR 72701 Email: dk030@uark.edu**

# Supplementary figures:

**S0: Mobility calculation**

We can estimate electron mobility in our device from the $I_d$-$V_{bg}$ trace shown below (same as in Figure 1I of the main text). The transconductance trace can be split into three regions. Below $V_{bg} < 6V$, the device is in the OFF state. Between 6V to 7V, $I_d$ increases linearly and above $V_{bg} > 7V$, the $I_d$ starts to saturate where the $I_d$ is limited by the contact resistance of the device. To calculate the mobility, we use the relation:

$$\mu = \frac{dG}{dV_{bg}} \frac{L}{W} \frac{1}{C_g} \quad \ldots (1)$$

Where $dG/dV_{bg}$ is the rate of change of conductnace as a funtion of backgate votlage, $L$ = length of the channel (10.25 μm), $W$ = width of the channel (8 μm), $G = I_d/V$ and $C_g$ is the gate capacitance. The gate capacitance can be calculated by considering the dielectric environment between MoS$_2$ and the backgate. In our case, the gate dielectric comprise of 285 nm of SiO$_2$ and 30 nm of hBN. Therefore the gate capacitance $C_g$ is given by:

$$C_g = \epsilon_0 \frac{\epsilon_{SiO2} \epsilon_{hBN}}{\epsilon_{hBN} t_{SiO2} + \epsilon_{SiO2} t_{hBN}}$$

$$C_g = \epsilon_0 \frac{(4)(4)}{4(285 \times 10^{-9}) + 4(30 \times 10^{-9})}$$

$$C_g = 1.12 \times 10^{-4} F/m^2 \quad \ldots (2)$$

Using eq.1, we fit the higher backgate voltage range to estimate a mobility of ~27,000 cm$^2$/V.s. However, we emphasize that it is difficult to extract the gate-channel capacitance accurately in dual gated device geometry using parallel plate capacitor model in two-terminal measurement configuration. Moreover, gate voltage dependent contact resistance needs to be taken into account to extract the exact mobiltiy. Hence, the actual mobility in our device can be upto one order of magnitude lower than the extracted value (i.e. few thousands cm$^2$/V.s.).

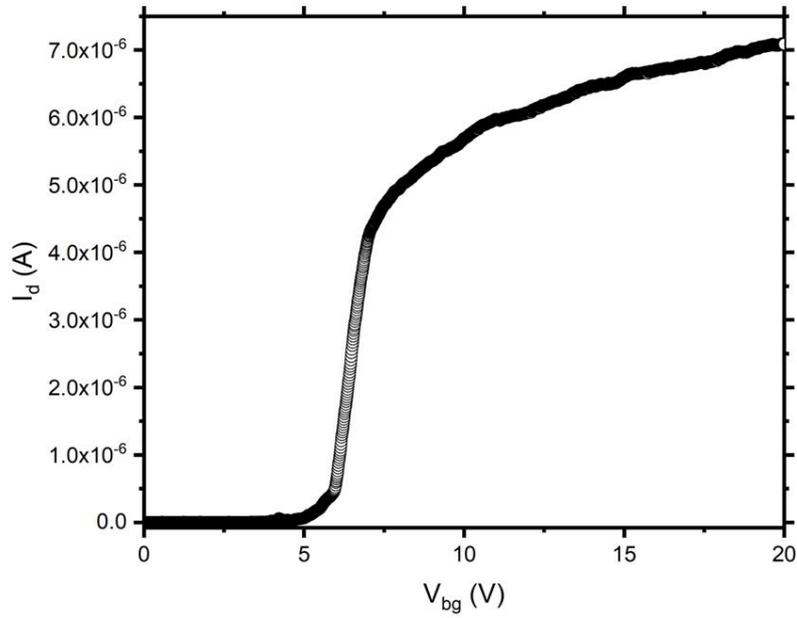

## S1: Zeeman splitting of crossing A shown in main text figure 2B

Evolution of CB diamond A shown in main text Figure 2B under perpendicular magnetic field. The figure below shows CB diamond A measured at $B_z = 0T$ (Figure S1A) and $B_z = 7.5T$ (Figure S1B). We see a spliting of ground state at higher magnetic field. From the splitting we extract a *g*-factor of 3.5 which is lower than the value observed for CB diamond B and C but still higher than the previous reports. The spin filling order in CB diamond A is consistent with that of CB diamond B shown in the main text.

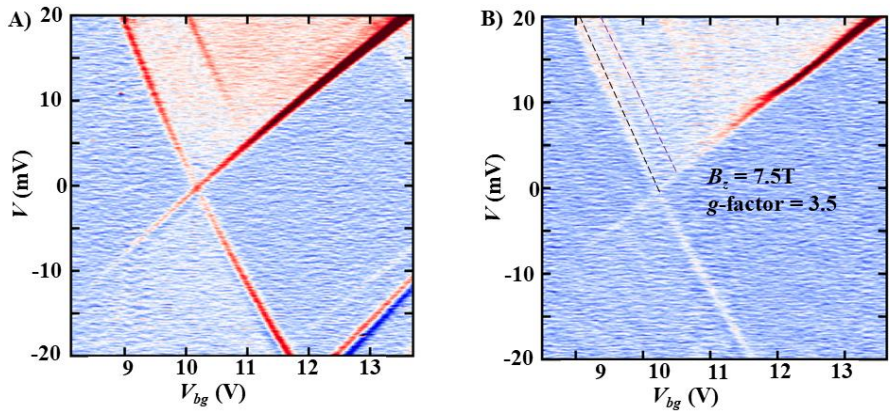

## S2: Zeeman splitting of crossing C shown in main text figure 2B

We measure the *g*-factor for CB diamond C as well. We compare the same CB diamond under different magnetic field to measure Zeeman splitting. Figure below B-E shows CB diamond B measured for $B_z$ = 0T, 2.5T, 5T and 7.5T. We observe Zeeman splitting of the ground state due to lifting of spin degeneracy (marked by yellow arrow in Figure S2 C-E below) which increases with increasing $B_z$. From the evolution of the Zeeman splitting as a function of the magnetic field, we can extract the electron $g$-factor for CB diamond C. In Figure F below we plot the extracted Zeeman energy as a function of the $B_z$ and we extract a $g$-factor of 5.20 ± 0.17. The spin filling order in CB diamond C is consistent with that of CB diamond B shown in the main text.

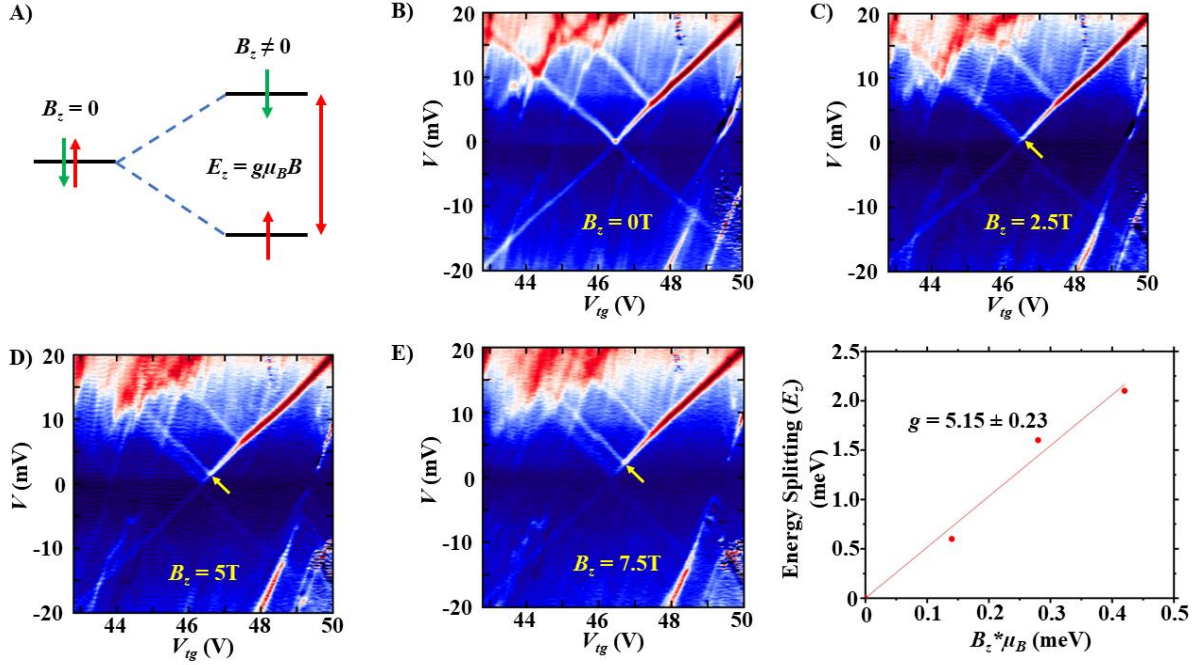

## S3: Dependence of quantum dot energy level on back gate and top gate

To examine the coupling of the top gate to the QDs, we measure the CB peaks as a function of $V_{bg}$ and $V_{tg}$ at $V$ = 5mV. For a QD equally coupled to both the gates, we expect a slope of -1 in the $V_{bg}$-$V_{tg}$ space. In our device, $V_{tg}$ is much closer to the $MoS_2$ channel (~ 10nm thick hBN dielectric) than $V_{bg}$ and hence an asymmetric coupling is expected. We look at the CB peak (Figure 5A) which corresponds to single charge transition and use the relation $Q = CV$ where Q =1e and C corresponds to capacitive coupling and V is the gate voltage. Since we are looking at single charge transition dependence as a function of $V_{tg}$ and $V_{bg}$, we can equate $V_{tg}/V_{bg} = C_{bg}/C_{tg}$.

where $C_{tg}$ is top gate capacitance coupling to QD and $C_{bg}$ is the back gate capacitance coupling to the QD. From the slope of the first charge transition in the figure below, we get a $C_{tg}$ = 69 * $C_{bg}$ supporting our claims that the QDs are strongly coupled and located below the top gate.

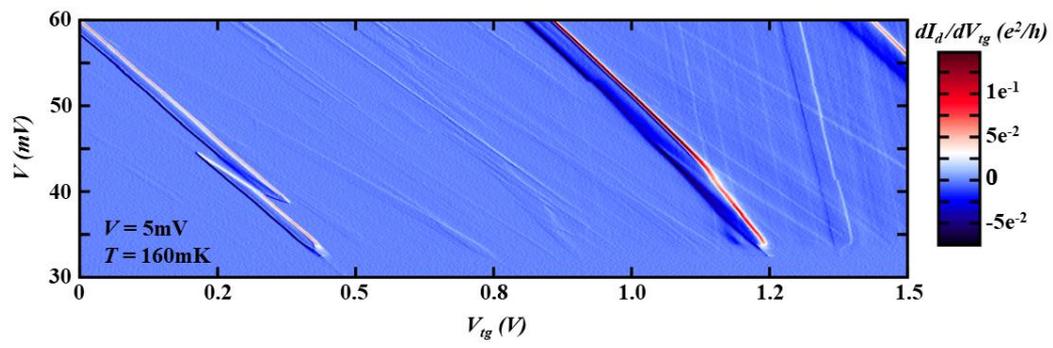